\documentclass[10pt,final,twocolumn]{IEEEtran}
\usepackage{mdwlist}
\usepackage{amssymb,amsfonts}
\usepackage[cmex10]{amsmath}
\usepackage{cite}

\newtheorem{lemma}{Lemma}
\newtheorem{theorem}{Theorem}
\newtheorem{proposition}{Proposition}

\newtheorem{corollary}{Corollary}
\newtheorem{definition}{Definition}
\newtheorem{remark}{Remark}
\newtheorem{assum}{Assumption}

\newtheorem{exam}{Example}

\def\vp{\varphi}
\def\ve{\varepsilon}

\def\om{\omega}

\def\r{\mathbb R}

\def\be{\beta}
\def\c{\mathbb C}

\def\S{\mathbb S}

\def\imath{\mathbf{i}}

\def\be{\begin{equation}}
\def\ee{\end{equation}}
\def\ben{\begin{equation*}}
\def\een{\end{equation*}}
\newcommand{\dfb}{\stackrel{\Delta}{=}}

\usepackage{graphicx,array,enumerate}
\usepackage{tikz}
\usetikzlibrary{calc,shapes,arrows,petri}
\tikzset{myrad/.style 2 args={circle,inner sep=0pt,minimum width=(2*(sqrt(#1)*1 cm ) - \pgflinewidth,fill=#2,draw=#2}}

\usepackage{caption,subcaption}

\begin{document}
\title{Synchronization of Pulse-Coupled Oscillators and Clocks under Minimal Connectivity Assumptions}

\author{Anton~V. Proskurnikov,~\IEEEmembership{Member,~IEEE}, Ming Cao,~\IEEEmembership{Senior Member,~IEEE}
\thanks{Anton V. Proskurnikov is with Delft Center for Systems and Control at Delft University of Technology, Delft, The Netherlands.
He is also with Institute for Problems of Mechanical Engineering of the Russian Academy of Sciences (IPME RAS) and ITMO University,
St. Petersburg, Russia, {e-mail: \tt anton.p.1982@ieee.org}}
\thanks{Ming Cao is with the ENTEG institute at the University of Groningen, Groningen, The Netherlands, {e-mail: \tt ming.cao@ieee.org}.}
\thanks{Partial funding was provided by the ERC (grant ERCStG-307207), STW (vidi-438730) and Russian Federation President's Grant MD-6325.2016.8. 
Theorem~2 was obtained under sole support of Russian Science Fund (RSF) grant 14-29-00142, hosted by IPME RAS.}
}

\maketitle

\begin{abstract}
Populations of flashing fireflies, claps of applauding audience, cells of cardiac and circadian pacemakers
reach synchrony via event-triggered interactions, referred to as \emph{pulse couplings}. Synchronization via pulse coupling is widely used in wireless sensor networks, providing clock synchronization with parsimonious packet exchanges. In spite of serious attention paid to networks of pulse coupled oscillators, there is a lack of mathematical results, addressing networks with general communication topologies and general phase-response curves of the oscillators. The most general results of this type (Wang et al., 2012, 2015) establish synchronization of oscillators with a delay-advance phase-response curve over strongly connected networks. In this paper we extend this result by relaxing the connectivity condition to the existence of a root node (or a directed spanning tree) in the graph. This condition is also necessary for synchronization.
\end{abstract}


\section{Introduction}

Recent development of hardware and software for computation and communication has opened up the possibility of large scale control systems, whose components
are spatially distributed over large areas. The necessity to use communication and energy-supply resources ``parsimoniously''
has given rise to rapidly growing theories of control under limited data-rate~\cite{MatveevSavkinBook} and event-triggered control~\cite{Tabuada:2007,MazoCao:2014}.
Many control and coordination algorithms, facing communication and computational constraints, have been inspired by natural phenomena, discovered long before the ``network boom'' in control. Early studies of the phenomenon of synchronous flashing in large populations of male fireflies in the dark~\cite{Buck:1938} have disclosed a vision-based distributed protocol,
enabling fireflies to synchronize their internal clocks: \emph{``each individual apparently took his cue to flash from his more immediate neighbors, so that the mass flash took the form of a very rapid chain of overlapping flashes...''} \cite[p. 310]{Buck:1938}.
In a similar way the claps of many hands synchronize into rhythmic applause \cite{NedaVicsek:2000}.
Later works revealed the role of such event-based interactions, referred to as the \emph{pulse coupling},
in synchronization of neural networks~\cite{IzhikevichBook}, in particular, the cells of cardiac~\cite{Peskin} and circadian~\cite{WinfreeBook} pacemakers.
Self-synchronizing networks of biological pulse-coupled oscillators (PCO) have inspired efficient algorithms for \emph{clock synchronization} in wireless networks~\cite{HongScaglione:2005,Pagliari:2011,WangDoyle:2012,WangNunezDoyle:2012,WangNunezDoyle:2013}, substantially reducing communication between the nodes.

The influential papers \cite{MirolloStrogatz:1990, Kuramoto:1991}, addressing the dynamics of PCO networks, attracted extensive attention from applied mathematicians, physicists and engineers, since ensembles of PCO give an instructive model of self-organization in complex systems, composed of very simple units. Each unit of the ensemble is a system, which operates in
a small vicinity of a stable \emph{limit cycle} and is naturally represented by a scalar \emph{phase} variable~\cite{SacreSepulchre:2014}.
An oscillator's phase varies in a bounded interval; upon achieving its maximum, the phase is reset to the minimal value.
At this time the oscillator fires an event, e.g. emitting electric pulse or other stimulus. The length of these pulses is usually neglected since they are very short, compared to the oscillators' periods. Unlike Kuramoto networks and other \emph{diffusively} coupled oscillator ensembles~\cite{BulloSurvey:2014,ProCao:2017-1}, the interactions of PCO are event-triggered. The effect of a stimulus from a neighboring oscillator on an oscillator's trajectory is modeled by a phase shift, characterized by the nonlinear \emph{phase response curve} (PRC) mapping \cite{Canavier:2010,SacreSepulchre:2014}.

In spite of significant interest in dynamics of PCO networks, the relevant mathematical results are very limited. Assuming that the oscillators are weakly coupled, the hybrid dynamics of PCO networks can be approximated by the Kuramoto model~\cite{Kuramoto:1991,WangDoyle:2012,WangNunezDoyle:2013} that has been thoroughly studied~\cite{BulloSurvey:2014}. The analytic results for general couplings are mostly confined to networks with special graphs~\cite{MirolloStrogatz:1990,GoelErmentrout:2002,LuckenYanchuk:2012,WangNunezDoyle:2015},
providing a fixed order of the oscillators' firing. In recent papers~\cite{WangNunezDoyle:2012,WangNunezDoyle:2015-1} synchronization criteria over general \emph{strongly connected} graphs have been obtained, assuming that oscillators' PRC maps are
\emph{delay-advance}~\cite{IzhikevichBook} and the deviations between the initial phases are less than a half of the oscillators' period.
The main idea of the proof in \cite{WangNunezDoyle:2012,WangNunezDoyle:2015-1} is the \emph{contracting property} of the network dynamics under the assumption of delay-advance PRC, enabling one to use the maximal distance between the phases (the ensemble's ``diameter'') as a Lyapunov function; this approach is widely used in the analysis of Kuramoto networks~\cite{SchmidtPapaAllgower:2012,Antonis}.

In this paper, we further develop the approach from~\cite{WangNunezDoyle:2012,WangNunezDoyle:2015-1}, relaxing the strong connectivity assumption to the existence of a directed spanning tree (or \emph{root} node) in the interaction graph, which is also necessary for synchronization. Also, unlike~\cite{WangDoyle:2012,WangNunezDoyle:2013} the delay-advance PRC maps are not restricted to be piecewise-linear and can be heterogeneous. Both extensions are important. Biological oscillator networks are usually ``densely'' connected (so the strong connectivity assumption is not very restrictive), but the piecewise linearity of PRC maps is an impractical condition. In clock synchronization problems the PRC map can be chosen piecewise-linear, but the requirement of strong connectivity excludes many natural communication graphs (e.g. the star-shaped graph with the single ``master'' clock and several ``slaves'').
The results have been partly reported in the conference paper~\cite{ProCao15}.

The paper is organized as follows. Preliminary Section~\ref{sec.prelim} introduces technical concepts and notation. The mathematical model of PCO networks is introduced in Section~\ref{sec.oscill}. Main results are formulated in Section~\ref{sec.main} and confirmed in Section~V by numerical simulations. Section~\ref{sec.concl} concludes the paper.

\section{Preliminaries and notation}\label{sec.prelim}

Given $t_0\in\r$ and a function $f(\cdot)$, defined at least on the interval $(t_0-\ve_0;t_0)$ for $\ve_0>0$ sufficiently small, let $f(t_0-)\dfb\lim\limits_{t\to t_0,t<t_0}f(t)$.
If $f(t_0-)=f(t_0)$, we say $f(\cdot)$ is \emph{left-continuous} at $t_0$. The limit $f(t_0+)$ and right-continuity are defined similarly.
A function $f:[0;+\infty)\to\r$ is \emph{piecewise continuous}, if it is continuous at any $t\ge 0$ except for a sequence $\{t_n\}_{n=1}^{\infty}$, such that $t_n\to\infty$ and at each of the points $t_n$ the left and right limits
$f(t_n-)$, $f(t_n+)$ exist.

We denote the unit circle on the complex plane by $\mathbb{S}^1=\{z\in\c:|z|=1\}$. Given $\vp\in\r$, $e^{\imath\vp}=\cos\vp+\imath\sin\vp\in\mathbb{S}^1$.
Here $\mathbf{i}$ stands for the imaginary unit, $\imath^2=-1$.

A (directed) \emph{graph} is a pair $(V,E)$, where $V$ and $E\subseteq V\times V$ are finite sets, whose elements are referred to as the \emph{nodes} and \emph{arcs} respectively.
A \emph{walk} in the graph is a sequence of nodes $v_1,v_2,\ldots,v_k$, where consecutive nodes are connected by arcs $(v_i,v_{i+1})\in E$.
A \emph{root} is a node, from which the walks to all other nodes exist. A graph having a root is called \emph{rooted} (this is equivalent to the existence of a \emph{directed spanning tree}); a graph in which any node is a root is called \emph{strongly connected}. 

\section{The problem setup}\label{sec.oscill}

An \emph{oscillator} with frequency $\omega>0$ (or, equivalently, period $T=2\pi/\omega$) is a dynamical system $\dot x(t)=f(x(t))$ with an exponentially stable $T$-periodic limit cycle $x^0(t)=x^0(t+T)$. Any solution $x(t)$, staying in the cycle's basin of attraction, converges as $t\to\infty$ to the function $x^0(\theta(t)/\omega)$.
Here $\theta(t)\in [0;2\pi)$ is a piecewise-linear function, referred to as \emph{phase} and treated as
``a normalized time, evolving on the unit circle''~\cite{SacreSepulchre:2014}. The phase grows linearly until it reaches $2\pi$ and then is reset:
\begin{gather}
\dot\theta(t)=\om\quad\text{while $\theta(t-)<2\pi$},\label{eq.freq-single}\\
\theta(t+)=0\quad\text{if $\theta(t-)=2\pi$}\label{eq.reset-single}.
\end{gather}

In this paper we deal with \emph{ensembles} of multiple oscillators~\eqref{eq.freq-single}, whose interactions are \emph{event triggered}.
Upon resetting, an oscillator fires an \emph{event} by sending out some \emph{stimulus} such as a short electric pulse or message.
If an oscillator receives a stimulus from one of its neighbors, its phase jumps
\be\label{eq.shift-single}
\theta(t+)=\Psi(\theta(t-))\mod 2\pi,\quad \Psi(\theta)\dfb\theta+c\Phi(\theta),
\ee
after which the ``free run''~\eqref{eq.freq-single} continues. Typically it is assumed that $\Phi(0)=\Phi(2\pi)=0$ so that if an oscillator is triggering an event at time $t$, then the stimuli received from the remaining oscillators
do not violate~\eqref{eq.reset-single}. The map $\Psi:[0;2\pi]\to\r$ is referred to as the oscillator's \emph{phase transition curve} (PTC)~\cite{IzhikevichBook}. The PTC is determined as in~\eqref{eq.shift-single} by the map $\Phi:[0;2\pi]\to \r$, referred to as the \emph{phase response (or resetting) curve} (PRC)~\cite{Canavier:2010,IzhikevichBook}, and the scalar \emph{coupling gain} $c>0$. In networks of biological oscillators, the PRC maps depend on the stimuli waveforms and the gain $c$ depends on the stimulus' intensity~\cite{BrownMoehlisHolmes:2004,IzhikevichBook,Canavier:2010,SacreSepulchre:2014}. 
In time synchronization problems~\cite{HongScaglione:2005,WangNunezDoyle:2012,WangNunezDoyle:2013} the PRC map $\Phi$ and the coupling gain $c$ are the
parameters to be designed.

Henceforth we assume\footnote{Dealing with ``weakly coupled'' PCO networks ($c\approx 0$)~\eqref{eq.shift-many-single} is often replaced by the additive rule $\theta(t+)=\theta(t)+kc\Phi(\theta(t))\mod 2\pi$, enabling one
to approximate the PCO network by the Kuramoto model~\cite{Kuramoto:1991}.}, following~\cite{LuckenYanchuk:2012}, that $k>1$ simultaneous events, affecting an oscillator, superpose as follows
\be\label{eq.shift-many-single}
\theta(t+)=\Psi^{k}(\theta(t-))\mod 2\pi,\; \Psi^{k}\dfb\underbrace{\Psi\circ\Psi\circ\ldots\circ\Psi}_{\text{$k$ times}}.
\ee
Taking $\Psi^0(\theta)\dfb\theta$,~\eqref{eq.shift-many-single} holds for $k=0$: if the neighbors fire no events, the phase is continuous unless
it has reached $2\pi$. Note that $\theta(t+)<2\pi$ at any point; in particular, the oscillator cannot be forced to fire due to its neighbors' stimuli.

At the points of discontinuity one can define $\theta(t)$ arbitrarily; for definiteness, we suppose that $\theta(t)=\theta(t-)\in [0;2\pi]$. We also allow the initial phase $\theta(0)=2\pi$: 
the oscillator fires an event and is immediately reset to $0$.

\subsection{Mathematical model of the PCO network}

 Consider a group of $N>1$ oscillators of the same period $T=2\pi/\omega$ and PTC mappings $\Psi_1(\theta),\ldots,\Psi_N(\theta)$, corresponding to PRC maps $\Phi_i$ and coupling gains $c_i>0$.
 The vector of oscillators' phases is denoted by $\bar\theta(t)\dfb(\theta_i(t))_{i=1}^N\in [0;2\pi]^N$.

The interactions among the oscillators are encoded by a graph $G=(V,E)$, whose nodes are in one-to-one correspondence with oscillators $V=\{1,\ldots,N\}$.
The arc $(j,i)$ exists if and only if oscillator $j$ influences oscillator $i$; we denote $N_i\dfb\{j:(j,i)\in E\}$ to denote the set of oscillators, affecting oscillator $i$;
it is convenient to assume that $i\in N_i\,\forall i$.

The dynamics of the PCO network is as follows
\begin{gather}
\dot{\bar\theta}(t)=(\omega,\ldots,\omega)\quad\text{when $I(\bar\theta(t))=\emptyset$,}\label{eq.freq}\\
\bar\theta(t+)=\bar\Psi(\bar\theta(t))\mod 2\pi\quad\text{if $I(\bar\theta(t))\ne\emptyset$}\label{eq.shift},\\
\bar\Psi(\theta_1,\ldots,\theta_N)\dfb \left(\Psi_1^{k_1}(\theta_1),\ldots,\Psi_N^{k_N}(\theta_N)\right)\label{eq.psi-bar},\\
I(\bar\theta)\dfb\{j:\theta_j=2\pi\},\quad k_i=k_i(\bar\theta)\dfb\left|I(\bar\theta)\cap N_i\right|\label{eq.indices}.
\end{gather}
Here $|\cdot|$ denotes the cardinality of a set.
The phases obey~\eqref{eq.freq-single} until some oscillators fire; $I(\bar\theta(t))\ne\emptyset$ stands for the set of their indices.
Oscillator $i$ is affected by $k_i\ge 0$ firing neighbors, and its phase jumps in accordance with~\eqref{eq.shift-many-single}. If $k_i=0$ then
$\theta_i(t)<2\pi$ (since $i\in N_i$) and $\theta_i(\cdot)$ is continuous at $t$.
\begin{definition}\label{def.solution}
A function $\bar\theta:\Delta\to [0;2\pi]^N$ is said to be a solution to the system~\eqref{eq.freq},~\eqref{eq.shift} on the interval $\Delta\subseteq [0;\infty)$ if the following conditions hold
\begin{enumerate}
\item on any \emph{compact} interval $\Delta'\subseteq \Delta$ only a finite number of events are fired $\left|\Delta'\cap\left\{t:I(\bar\theta(t))\ne\emptyset\right\}\right|<\infty$;
\item the function $\bar\theta(t)$ is left-continuous and obeys~\eqref{eq.freq} at any $t\ge 0$ except for the points where some oscillators fire; at such points $\bar\theta(t)$ switches in
accordance with~\eqref{eq.shift}.
\end{enumerate}
\end{definition}
\begin{remark}\label{rem.definitions-difference}
Our definition of a solution is more restrictive than the definitions in~\cite{WangNunezDoyle:2015,WangNunezDoyle:2015-1}, which replace the discontinuous mapping $\bar\Psi$ in~\eqref{eq.shift} by an outer-semicontinuous~\cite{Goebel:2009} \emph{multi-valued} map. Unlike the ``generalized'' solutions from~\cite{WangNunezDoyle:2015,WangNunezDoyle:2015-1}, the solution from Definition~\ref{def.solution} is \emph{uniquely} determined by its initial condition $\bar\theta(0)$ and depends continuously on it.
\end{remark}

Our goal is to establish conditions, under which the solution $\bar\theta(t)$ to the system~\eqref{eq.freq},~\eqref{eq.shift} exists on $[0;\infty)$ and the oscillators' phases
become~\emph{synchronous} in the following sense.
\begin{definition}\label{def.sync}
The phases $\theta_i(\cdot)$ ($i\in 1:N$) synchronize if
\be\label{eq.sync}
e^{\imath(\theta_i(t)-\theta_j(t))}\xrightarrow[t\to\infty]{} 1\Longleftrightarrow e^{\imath\theta_i(t)}-e^{\imath\theta_j(t)}\xrightarrow[t\to\infty]{} 0.
\ee
\end{definition}

\subsection{Assumptions}

In this subsection, we formulate two assumptions adopted throughout the paper. The first of these assumptions implies an important contraction property of the hybrid dynamics~\eqref{eq.freq},\eqref{eq.shift}.
\begin{assum}\label{ass.psi}
The mappings $\Psi_i$ are continuous on $[0;2\pi]\setminus\{\pi\}$, satisfying the conditions $\Psi_i(0)=0$, $\Psi_i(2\pi)=2\pi$ and
\ben
\Psi_i(\theta)\in (0;\theta)\,\forall \theta\in (0;\pi),\quad \Psi_i(\theta)\in(\theta;2\pi)\,\forall \theta\in  (\pi;2\pi).
\een
\end{assum}
\begin{figure}
\center
\begin{subfigure}[t]{0.4\linewidth}
\begin{tikzpicture}[scale=0.4, baseline=(A.base)]
\draw (-2.5,0) -- (2.5,0);
\draw (0,-2.5) -- (0,2.5);
\draw (0,0) circle (2cm);
\draw (0,2) [red, very thick] arc (90:30:2cm);
\draw (0,2) [->,red, very thick] arc (90:60:2cm);
\node [label=above right:\small $\theta_i(t)$]  at (0,2) [line width=0.05mm,myrad={0.02}{black}] {};
\node [label=right:\color{blue}\small $\theta_i(t+)$]  at (1.732,1) [line width=0.05mm,myrad={0.02}{blue}] {};
\node [label=right:{\small $2\pi=\theta_{j}(t)$} ]  (A) at (2,0) [line width=0.05mm,myrad={0.02}{gray}] {};
\end{tikzpicture}
\end{subfigure}
\begin{subfigure}[t]{0.4\linewidth}
\begin{tikzpicture}[scale=0.4, baseline=(A.base)]
\draw (-2.5,0) -- (2.5,0);
\draw (0,-2.5) -- (0,2.5);
\draw (0,0) circle (2cm);
\draw (0,-2) [red, very thick] arc (-90:-45:2cm);
\draw (0,-2) [->,red, very thick] arc (-90:-60:2cm);
\node [label=below right :\small $\theta_i(t)$]  at (0,-2) [line width=0.05mm,myrad={0.02}{black}] {};
\node [label=right:\color{blue}\small $\theta_i(t+)$]  at (1.732,-1) [line width=0.05mm,myrad={0.02}{blue}] {};
\node [label=right:{\small $2\pi=\theta_{j}(t)$} ] (A) at (2,0) [line width=0.05mm,myrad={0.02}{gray}] {};
\end{tikzpicture}
\end{subfigure}
\caption{Illustration to Assumption~\ref{ass.psi}: the jump~\eqref{eq.shift-single} decreases the distance between the oscillator $i$ and its firing neighbor $j$.}\label{fig.d1}
\end{figure}
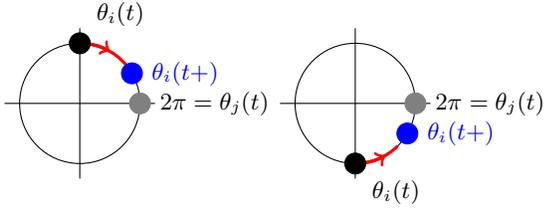
Assumption~\ref{ass.psi} is illustrated by~Fig.~\ref{fig.d1}. The $i$th ``clock'' is \emph{delayed} by the phase jump~\eqref{eq.shift-many-single}
if it is ahead of its firing neighbors (Fig.~\ref{fig.d1}, left part) and \emph{advanced} if it is behind them (Fig.~\ref{fig.d1}, right part).
Such operations do not lead to ``overshoots'': a ``retarding'' oscillator cannot overrun its neighbors and become ``advancing'', and vice versa.
A firing oscillator is not influenced by the others' events since $\Psi_i^{k_i}(2\pi)\mod 2\pi=0$.

Assumption~\ref{ass.psi} holds, in particular, for PCOs with coupling gains $c_i\in (0;1)$ and piecewise-linear PRC maps
\be\label{eq.prc-lin}
\Phi_1(\theta)=\ldots=\Phi_N(\theta)=\begin{cases}
-\theta,&\theta\in [0;\pi)\\
2\pi-\theta,&\theta\in (\pi;2\pi]\\
\text{any}, &\theta=\pi.
\end{cases}
\ee
Such a choice of the PRC map appears to be the most natural in time synchronization problems~\cite{WangDoyle:2012,WangNunezDoyle:2013,WangNunezDoyle:2015,WangNunezDoyle:2015-1}.
More generally, the PRC map $\Phi(\theta)$ is called \emph{delay-advance}~\cite{WangNunezDoyle:2012} if $\Phi(\theta)<0$ for $\theta\in (0;\pi)$ and $\Phi(\theta)>0$ when $\theta\in(\pi;2\pi)$.
Mathematical models of natural oscillators with delay-advance PRC include, but are not limited to, ``isochron clocks''~\cite{GoelErmentrout:2002} and the Andronov-Hopf oscillator~\cite{IzhikevichBook}. Assumption~\ref{ass.psi} holds for sufficiently small $c_i>0$ if $\Phi_i$ are delay-advance and
\ben
\inf_{\theta\in (0;\pi)}\frac{\Phi_i(\theta)}{\theta}>-\infty\quad\text{and}\quad \sup_{\theta\in (\pi;2\pi)}\frac{\Phi_i(\theta)}{2\pi-\theta}<\infty\quad\forall i.
\een

To introduce our second assumption, restricting oscillators to be ``partially synchronous'', we need a technical definition.
\begin{definition}\label{def.diam}
An \emph{arc} of $\S^1$ is a closed connected subset $L\subseteq\S^1$. Given a vector of phases $\bar\theta=(\theta_i)_{i=1}^N$, its \emph{diameter} $d(\bar\theta)$ is the length of the shortest arc, containing the set $\{e^{\imath\theta_i}\}_{i=1}^N$.
\end{definition}

\begin{figure}
\center
\begin{tikzpicture}[scale=0.4]
\draw (-2.5,0) -- (2.5,0);
\draw (0,-2.5) -- (0,2.5);
\draw (0,0) circle (2cm);
\draw (0,-2) [red, very thick] arc (-90:135:2cm);
\node [label=above:\small $\theta_1$]  at (-1.414,1.414) [line width=0.05mm,myrad={0.02}{black}] {};
\node [label=above:\small $\theta_2$]  at (1.414,1.414) [line width=0.05mm,myrad={0.02}{black}] {};
\node [label=below right:\small $\theta_3$]  at (0,-2) [line width=0.05mm,myrad={0.02}{black}] {};
\end{tikzpicture}
\caption{$\bar\theta=(\pi/4,3\pi/4,3\pi/2)$, $d(\bar\theta)=5\pi/4$}\label{fig.d}
\end{figure}
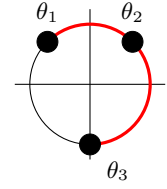

The definition of diameter is illustrated by Fig.~\ref{fig.d}: one of the two shortest arcs, containing the phases, is drawn in red.

\begin{assum}\label{ass.initial}
The initial phases of the oscillators are ``partially synchronized'', satisfying the inequality
\be\label{eq.initial}
d(\bar\theta(0))<\pi.
\ee
\end{assum}

\begin{remark}\label{rem.no-synchronize}
The ``partial synchronization'' Assumption~\ref{ass.initial} can be relaxed in some special situations~\cite{MirolloStrogatz:1990,GoelErmentrout:2002}, but generally cannot be fully discarded. The simplest example is a network of $N=2$ coupled oscillators, whose PRC maps $\Phi_1,\Phi_2$ satisfy the condition $\Phi_1(\pi)=\Phi_2(\pi)=0$. Then the solution, starting at $(\theta_1(0),\theta_2(0))=(0;\pi)$, is $T$-periodic and $d(\bar\theta(t))\equiv \pi$.
Conditions similar to~\eqref{eq.initial} are often adopted to prove the synchronization of diffusively coupled oscillators~\cite{BulloSurvey:2014}.
\end{remark}

\section{Main result}\label{sec.main}

We start with establishing basic properties of the dynamical network~\eqref{eq.freq},~\eqref{eq.shift} (Subsect.~\ref{subsec.basic})
and then prove the the main result of the paper, ensuring synchronization (Subsect.~\ref{subsec.synch}).

Our method extends the idea of the diameter Lyapunov function, used to prove stability of multi-agent coordination protocols~\cite{Moro:05}, to the hybrid system~\eqref{eq.freq},~\eqref{eq.shift}. We show that the diameter $d(\bar\theta(t))$ of the oscillator ensemble is non-increasing and, furthermore, there exists a period $T_N$, independent of the initial condition, such that $d(\bar\theta(T_N))-d(\bar\theta(0))<0$ unless $d(\bar\theta(0))=0$.
The key idea is to establish the LaSalle-type result for the hybrid system~\eqref{eq.freq},~\eqref{eq.shift} and the Lyapunov function $d(\cdot)$, stating that any solution converges to the synchronous manifold $\{\bar\theta\in [0;2\pi]^N:d(\bar\theta)=0\}$. In the existing literature~\cite{WangNunezDoyle:2012,WangNunezDoyle:2015-1}, this is
done via a straightforward estimation of the diameter's decrease $d(\bar\theta(T_N))-d(\bar\theta(0))$, employing the special structure of PRC maps and the strong connectivity of the graph. We extend these results to the case of rooted graphs and general delay-advanced PRC maps, deriving the mentioned LaSalle-type result from the continuity of the trajectory with respect to the initial condition.

\subsection{Basic properties of the solutions}\label{subsec.basic}

We first show existence and uniqueness of solutions to the system~\eqref{eq.freq},~\eqref{eq.shift} and establish their basic properties.
\begin{theorem}\label{thm.exist}
Under Assumption~\ref{ass.psi}, for \emph{any} initial condition $\bar\theta(0)\in [0;2\pi]^N$ the following statements hold:
\begin{enumerate}
\item the solution to~\eqref{eq.freq},~\eqref{eq.shift} exists on $[0;\infty)$ and is unique;
\item if some oscillator fires two consecutive events at instants $t'>0$ and $t''>t'$ respectively, then $t''-t'>T/2$;
\suspend{enumerate}
If the initial condition satisfies the inequality~\eqref{eq.initial}, then
\resume{enumerate}
\item the diameter function $d(t)\dfb d(\bar\theta(t))$ is non-increasing;
\item let $L(t)=L(\bar\theta(t))$ be the arc of the minimal length, containing $\{e^{\imath\theta_j(t)}\}_{j=1}^N$, then $L(t)\subseteq e^{\imath\om (t-t_0)}L(t_0)$
and $L(t_0+)\subseteq L(t_0)$ whenever $t>t_0\ge 0$;
\item for any $s\ge 0$ each oscillator fires on $(s;s+3T/2)$.
\end{enumerate}
\end{theorem}
\begin{remark}
The problem of solution existence has been studied in~\cite{WangNunezDoyle:2015} (Proposition~4) and~\cite{WangNunezDoyle:2015-1} (Proposition~1), using the general framework of
hybrid systems theory~\cite{Goebel:2009}. However, as discussed in Remark~\ref{rem.definitions-difference}, these results do not imply the existence of solutions in the sense of Definition~1.
The proofs of Theorem~1 in~\cite{WangNunezDoyle:2012} and Theorem~1 in \cite{WangNunezDoyle:2015-1}
contain in fact statements~3) and 4) for special PRC maps~\eqref{eq.prc-lin}. However, the proof of Theorem~\ref{thm.exist} for general delay-advance oscillators seems not to be available in the literature.
\end{remark}

The proof of Theorem~\ref{thm.exist} relies on the following proposition, proved in Appendix A.
\begin{proposition}\label{prop.uniqueness}
For a vector $\bar\xi\in [0;2\pi]^N$, denote
\[
\bar\xi^+\dfb\bar\Psi(\bar\xi)\mod 2\pi,\quad \delta_0\dfb T-\om^{-1}\max_i\xi_i^+>0.
\]
Then on the interval $\Delta_0=[0;\delta_0)$ the system~\eqref{eq.freq},~\eqref{eq.shift} has a unique solution with the initial condition $\bar\theta(0)=\bar\xi$. On $(0;\delta_0)$ no events are fired (events at time $t=0$ are possible).
\end{proposition}

\begin{figure}
\center
\begin{subfigure}[t]{0.4\linewidth}
\begin{tikzpicture}[scale=0.4, baseline=(A.base)]
\draw (-2.5,0) -- (2.5,0);
\draw (0,-2.5) -- (0,2.5);
\draw (0,0) circle (2cm);
\draw (0,2) [red, very thick] arc (90:-30:2cm);
\node [label=above right:\small $\theta_1(s)$]  at (0,2) [line width=0.05mm,myrad={0.02}{black}] {};
\node [label=right:\color{blue}\small $\theta_2(s)$]  at (1.732,1) [line width=0.05mm,myrad={0.02}{blue}] {};
\node [label=right:{\small $\theta_3(s)$} ]  at (1.732,-1) [line width=0.05mm,myrad={0.02}{gray}] {};
\node (A) at (2,0) {};
\node [label=below:\small $L(s)$] at (0,-2) {};
\end{tikzpicture}
\end{subfigure}
\begin{subfigure}[t]{0.4\linewidth}
\begin{tikzpicture}[scale=0.4, baseline=(A.base)]
\draw (-2.5,0) -- (2.5,0);
\draw (0,-2.5) -- (0,2.5);
\draw (0,0) circle (2cm);
\draw (-1.732,-1) [red, very thick] arc (-150:-30:2cm);
\node at (1.732,-1) [line width=0.05mm,myrad={0.02}{black}] {};
\node at (0,-2) [line width=0.05mm,myrad={0.02}{blue}] {};
\node at (-1.732,-1) [line width=0.05mm,myrad={0.02}{gray}] {};
\node (A) at (2,0) {};
\node [label=below:{\small $e^{\imath\om\tau}L(s),\;\tau\in (0;T)$}] at (0,-2) {};
\end{tikzpicture}
\end{subfigure}
\caption{Illustration to the proof of statement 4): rotation by some angle $\om\tau\in(0;2\pi)$ brings $L(s)$ to the lower half-plane.}\label{fig.rot}
\end{figure}
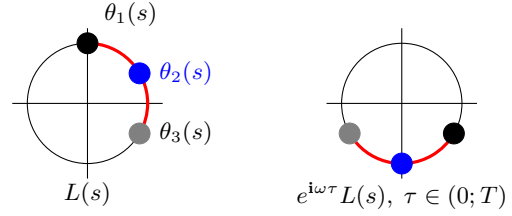

\begin{IEEEproof}[Proof of Theorem~\ref{thm.exist}]
We start with proving the implication: if the system has a solution (defined on some interval)
then for this solution statement 2) holds.
We are going to prove a more general fact: if a solution $\bar\theta(\cdot)$ exists on $[t';t]$, where $t'<t$ and $0\le\theta_i(t'+)\le\pi-\om(t-t')$ for some $i$,
then
\be\label{eq.aux1}
0<\theta_i(t)\le \theta_i(t'+)+\om(t-t')\le\pi.
\ee
In particular, if $\theta_i(t'+)=0$ and $t-t'\le T/2$, then $\theta_i(t)\le\pi$ and thus oscillator $i$ cannot fire at time $t$.
To prove~\eqref{eq.aux1}, recall that by Definition~\ref{def.solution} only a \emph{finite} number of events are fired between $t'$ and $t$. Denote the corresponding instants
$t_1<\ldots<t_n$. Since $\theta_i(t_1)=\theta_i(t'+)+\om(t_1-t')\in [0;\pi)$ and thus
$0\le\theta_i(t_1+)\le\theta_i(t_1)$. Iterating this procedure for $t_2,\ldots,t_n$,
one shows that $0\le\theta_i(t_n+)\le \om(t_n-t')+\theta_i(t'+)\le\pi$, which entails~\eqref{eq.aux1} since
$\theta_i(t)=\theta_i(t_n+)+\om(t-t_n)$.

To prove statement~1), we invoke Proposition~\ref{prop.uniqueness}, showing that
the solution exists and is unique on $\Delta=[0;\delta)$ for $\delta>0$ is sufficiently small.
Consider the \emph{maximal} interval $\Delta=[0;\delta)$ with this property. We are going to show that $\delta=\infty$.
Suppose on the contrary that $\delta<\infty$. Statement 2) shows that each oscillator fires a \emph{finite} number of events (at most $\lceil 2\delta/T\rceil$) on $\Delta$. Denoting the \emph{last} event instant by $t_*<\delta$, the phases obey~\eqref{eq.freq} on $(t_*,\delta)$ and hence the limit $\bar\theta(\delta)\dfb\bar\theta(\delta-)\in [0;2\pi]^N$ is defined. Applying Proposition~\ref{prop.uniqueness} to $\bar\xi\dfb \bar\theta(\delta)$, the solution is prolonged uniquely to $[\delta;\delta+\ve)$ for small $\ve>0$ and one arrives at a contradiction. Statement 1) is proved.

Statements 3) and 4) are proved analogously to the inequality~\eqref{eq.aux1}.  If $d(t)<\pi$ at the instant when some oscillators fire, then $L(t+)\subseteq L(t)$ and thus $d(t+)\le d(t)$ thanks to Assumption~\ref{ass.psi} since the new phases $\theta_i(t+)$ belongs to $L(t)$ (see Fig.~\ref{fig.d1}). Considering any interval $[t';t]$ (where $t'<t$) and the instants of events
$t_1<\ldots<t_n\le t$, one has
\be\label{eq.aux0}
\begin{split}
L(t)=e^{\imath\om(t-t_n)}L(t_n+)\subseteq e^{\imath\om(t-t_n)}L(t_n)\subseteq \\ \subseteq e^{\imath\om(t-t_{n-1})}L(t_{n-1})\subseteq\ldots\subseteq e^{\imath\om(t-t_{1})}L(t_{1})\subseteq\\\subseteq e^{\imath\om(t-t')}L(t').
\end{split}
\ee
It remains to prove statement 5). Retracing the proof of~\eqref{eq.aux1}, one proves that if $\theta_i(s)\in (\pi;2\pi)\,\forall s\in [t',t)$ then
\be\label{eq.aux2}
\theta_i(t)\ge \theta_i(t')+\om(t-t').
\ee
Hence if $\theta_i(t')\in (\pi;2\pi)$, oscillator $i$ fires on $(t';t'+T/2)$.
For any $s\ge 0$ there exists such $\tau\in [0;T)$ that $L(s+\tau)=e^{\imath\om\tau}L(s+)\subseteq \{e^{\imath\theta}:\theta\in (\pi;2\pi)\}$ (Fig.~\ref{fig.rot}).
Thus $\theta_i(s+\tau)\in (\pi;2\pi)$ for any $i$, and therefore each oscillator fires during the interval $(s+\tau;s+\tau+T/2)\subseteq (s;s+3T/2)$.
\end{IEEEproof}

\begin{remark}
Statements 2) and 5) of Theorem~\ref{thm.exist} show that under Assumptions~\ref{ass.psi} and~\ref{ass.initial} the time elapsed between two \emph{consecutive} events, fired by the same oscillator,
lies strictly between $T/2$ and $3T/2$. Both bounds are tight and cannot be relaxed, as demonstrated by the following example.
\end{remark}
\begin{exam}
Consider a network of two oscillators ($N=2$) with $T=2\pi$, PRC map~\eqref{eq.prc-lin} and gain $c\in(0;1)$, whose graph contains the only arc $2\mapsto 1$.
Starting at $\theta_1(0)=0$ and $\theta_2(0)=\theta_*>\pi$, oscillator $2$ fires at time $t_*\dfb2\pi-\theta_*<\pi$ and $\theta_1(t_*+)=(1-c)t_*$.
Hence the next event is fired by oscillator $1$ at time $t_*+2\pi-(1-c)t_*=2\pi+c(2\pi-\theta_*)$.
If $\theta_*<\pi$, then $t_*>\pi$ and $\theta_1(t_*+)=(1-c)t_*+c2\pi$. Thus oscillator $1$ fires the next event at $t=t_*+(1-c)\theta_*=2\pi-c\theta_*$.
When $c\approx 1$ and $\theta_*\approx\pi$ the time elapsed between two events of oscillator $1$ can be arbitrarily close to both $T/2$ and $3T/2$.
\end{exam}

Henceforth we confine ourselves to the trajectories satisfying Assumption~\ref{ass.initial}. It appears that such trajectories \emph{continuously} depend on the initial conditions in the following sense. For a given solution $\bar\theta(t)$, let $t_{ik}=t_{ik}[\bar\theta(0)]$ stand for the time instant when oscillator $i$ fires its $k$th event.
\begin{lemma}\label{lem.contin}
Suppose that Assumption~\ref{ass.psi} holds. Consider a sequence of solutions $\bar\theta^{(n)}(t)$ such that $\bar\theta^{(n)}(0)\xrightarrow[n\to\infty]{}\bar\theta(0)$, where $d[\bar\theta(0)]<\pi$. Then $t^{(n)}_{ik}\dfb t_{ik}[\bar\theta^{(n)}(0)]\xrightarrow[n\to\infty]{} t_{ik}$. Furthermore, $\bar\theta^{(n)}(t)\xrightarrow[n\to\infty]{}\bar\theta(t)$ whenever $t\ne t_{ik}\,\forall i,k$.
\end{lemma}

To prove Lemma~\ref{lem.contin}, we use a technical proposition, which is based on Assumption~\ref{ass.psi} and proved in Appendix.
\begin{proposition}\label{prop.positive-time}
For any $d_*<\pi$ and $\delta>0$ there exists $\tau=\tau(d_*,\delta)>0$ such that if $\theta_i(0)\le 2\pi-\delta$ and $d(\bar\theta(0))<d_*$, then oscillator $i$ fires at no earlier than $t=\tau$
(i.e. $t_{i1}\ge\tau$).
\end{proposition}

Proposition~\ref{prop.positive-time} has the following corollary, entailing that the ``leading'' oscillators, whose initial phases are sufficiently close to the maximal one,
fire earlier than the remaining oscillators.
\begin{corollary}\label{cor.fire-first}
For any $d_*<\pi$, $\delta>0$ there exists $\ve=\ve(\delta,d_*)>0$ with the following property: for the phases satisfying the condition $\theta_1(0)\ge\ldots\ge\theta_m(0)>\theta_1(0)-\ve$ and $\theta_{m+1}(0),\ldots,\theta_N(0)\le \theta_1(0)-\delta$, oscillators $1,\ldots,m$ fire earlier than the remaining ones; moreover,
$t_*\le t_{i1}<t_*+\om^{-1}\ve<t_{j1}$ whenever $1\le i\le m<j\le N$. Here $t_*\dfb T-\om^{-1}\theta_1(0)$ stands for the instant of first event.
\end{corollary}
\begin{IEEEproof}
Obviously, oscillator $1$ fires at time $t_{11}=t_*$, and the phases obey~\eqref{eq.freq} until its event. Since $\theta_j(t_*)\le 2\pi-\delta$ for $j>m$, one has
$t_{j1}>t_*+\tau$, where $\tau=\tau(\delta,d_*)$ is defined in Proposition~\ref{prop.positive-time}. Choosing $\ve<\min(\om\tau,\pi)$, one notices that $\theta_j(t_*)\ge 2\pi-\ve$ for $i=1,\ldots,m$.  Using~\eqref{eq.aux2}, oscillator $i$ fires at time $t_{i1}\le t_*+\om^{-1}\ve<t_{j1}$, which ends the proof.
\end{IEEEproof}

We are now ready to prove Lemma~\ref{lem.contin}.

\begin{IEEEproof}[Proof of Lemma~\ref{lem.contin}]
For the solution $\bar\theta(t)$, let $\tau_1<\tau_2<\ldots<\tau_n$ be the instants when some oscillators fire, i.e. $I_j\dfb I(\bar\theta(\tau_j))\ne\emptyset$.
Without loss of generality, one may assume that $I_1=\{1,\ldots,m\}$, i.e. $\theta_1(0)=\ldots=\theta_m(0)>\theta_j(0)$ for any $j>m$.
Notice first that $\bar\theta^{(n)}(t)\xrightarrow[n\to\infty]{}\bar\theta(t)$ for any $t\in [0;\tau_1)$. Indeed, $\tau_1=T-\om^{-1}\theta_1(0)>t$ implies that $T-\om^{-1}\max_i\theta_i^{(n)}(0)>t$ for large $n$, and hence $\theta_i^{(n)}(t)=\theta_i^{(n)}(0)+\om t\to \theta_i(0)+\om t=\theta_i(t)$ for any $i$.

Applying Corollary~\ref{cor.fire-first}, one proves that $t_{i1}^{(n)}\to \tau_1$ for any $i\le m$ and $t_*^{(n)}\dfb\max_{i\le m}t_{i1}^{(n)}<\min_{j>m}t_{j1}^{(n)}$.
Using~\eqref{eq.aux1}, one shows that $0\le \theta_i^{(n)}(t_*^{(n)}+)\le \om(t_*^{(n)}-t_{i1}^{(n)})\to 0=\theta_i(\tau_1+)$ for any $i\le m$.
The same holds for the remaining phases $\theta_j^{(n)}(t_*^{(n)}+)\to\theta_j(\tau_1+)$ (where $j>m$) since the cumulative effect of $m$ events, separated by infinitesimally small time periods,
is the same as that of $m$ simultaneous events. Thus we have proved that $\bar\theta^{(n)}(t_*^{(n)}+)\xrightarrow[n\to\infty]{} \bar\theta(\tau_1+)$.

We now can iterate this procedure, replacing $\bar\theta(0)$ and $\bar\theta^{(n)}(0)$ with, respectively, $\bar\theta(\tau_1+)$ and $\bar\theta^{(n)}(t_*^{(n)}+)$. One shows that
 $\bar\theta^{(n)}(t)\xrightarrow[n\to\infty]{}\bar\theta(t)$ for any $t\in (\tau_1,\tau_2)$ and for large $n$ the group of oscillators with indices from $I_2$ fires their events at times converging to $\tau_2$. The value of the $n$th state $\bar\theta^{(n)}$ after the last of these events converges to $\bar\theta(\tau_2+)$, and so on.
\end{IEEEproof}

\subsection{Synchronization criterion}\label{subsec.synch}

Up to now, we have not assumed any connectivity properties, required to provide the oscillators' synchronization. The \emph{minimal} assumption of this type is the existence of a root (or, equivalently, a directed spanning tree) in the interaction graph $G$. In a graph without roots there exist two non-empty subsets of nodes, which have no incoming arcs and thus are ``isolated'' from each other and the remaining graph~\cite[Theorem 5]{Moro:05}. Obviously, the corresponding two groups of oscillators are totally independent of each other and thus do not synchronize.

The following theorem shows that under Assumptions~\ref{ass.psi} and~\ref{ass.initial} rootedness is sufficient for the synchronization~\eqref{eq.sync}.
\begin{theorem}\label{thm.fix}
Suppose that Assumptions~\ref{ass.psi} and~\ref{ass.initial} hold, and the interaction graph $G$ is \emph{rooted}. Then the phases get synchronous~\eqref{eq.sync}.
\end{theorem}

For \emph{strongly} connected interaction graphs and special PRC maps Theorem~\ref{thm.fix} has been established in \cite{WangNunezDoyle:2012,WangNunezDoyle:2015-1}.
The fundamental property of the dynamics~\eqref{eq.freq},~\eqref{eq.shift} (see the proofs of Theorem~1 in~\cite{WangNunezDoyle:2012} and Theorem~1 in~\cite{WangNunezDoyle:2015-1}) is ``contraction''
of the minimal arc, containing the phases, after each ``full round'' of the oscillators' firing. As soon as each of the $N$ oscillators has fired (some of them can fire twice),
the diameter of the ensemble is decreased. This property, however, \emph{does not} hold for a general rooted graph, as shown by the following.

\emph{Example 2.} Consider $N=3$ oscillators with the period $T=2\pi$rad/s that are connected in a chain $1\mapsto 2\mapsto 3$; thus $1$ is a root node, yet the graph is not strongly connected. Suppose that the oscillators start with $\theta_1(0)=0$, $\theta_2(0)=\theta_3(0)=\theta_0<\pi$. The events fired by oscillators $2$ and $3$ at the instant $t_1=2\pi-\theta_0$ do not affect oscillator $1$, and hence $\theta_1(t_1+)=\theta_1(t_1)=2\pi-\theta_0$. The latter oscillator fires at time $t_2=2\pi$ after which one has $\theta_1(t_2+)=0$, $\theta_2(t_2+)=\Psi(\theta_0)\in (0;\theta_0)$ and
$\theta_3(t_2+)=\theta_0$. Thus after the full round of firing the diameter remains equal to $\theta_0$. Considering a similar chain of $N>3$ oscillators, its diameter in fact may remain unchanged even after $(N-2)$ full rounds of firing (each oscillator has fired at least $N-2$ times).

It appears, however, that after $N-1$ ``full rounds'' of firing the diameter always decreases, which is the key idea of the proof of Theorem~\ref{thm.fix}.
\begin{lemma}\label{lem.shrink}
Under the assumptions of Theorem~\ref{thm.fix}, let $T_N\dfb 3T(N-1)/2$ and thus on $[0;T_N]$ each oscillator fires at least $(N-1)$ events. Then $d(
\bar\theta(T_N))<d(\bar\theta(0))$ unless $d(\bar\theta(0))=0$.
\end{lemma}
\begin{IEEEproof}
Introducing the shortest arc $L(t)$ from Theorem~\ref{thm.exist}, consider the sets of its endpoints
$\{e^{\imath\theta_j(t)}:j\in J_-(t)\}$ and $\{e^{\imath\theta_j(t)}:j\in J_+(t)\}$. The shortest turn from the phases, indexed by $J_-(t)$, to those indexed by $J_+(t)$ is counterclockwise, see Fig.\ref{fig.j}. A closer look at the proof of statements 2 and 3 in Theorem~\ref{thm.exist} reveals that at any time $t_*$, when some oscillators fire, the following alternatives are possible:
\begin{enumerate}[A)]
\item none of the ``extremal'' oscillators from $J_-(t_*)\cup J_+(t_*)$ is affected by the events; in this case $J_-(t_*+)=J_-(t_*)$, $J_+(t_*+)=J_+(t_*)$ and $d(t_*+)=d(t_*)$;
\item some of the ``extremal'' oscillators are affected, however $d(t_*+)=d(t_*)$; this implies that $J_-(t_*+)\subseteq J_-(t_*)$, $J_+(t_*+)\subseteq J_+(t_*)$ and one of these inclusions is strict;
\item some of the ``extremal'' oscillators are affected, and the diameter is decreased: $d(t_*+)<d(t_*-)$.
\end{enumerate}

Notice that during the ``full round'' of events (each oscillator fires at least once) the second or third must take place. Indeed, suppose that $J_-$ and $J_+$ remain constant during such a round. The graph's rootedness implies~\cite[Theorem 5]{Moro:05} that at least one of the corresponding sets of nodes has an arc, coming from outside. That is, a node $j\in J_-$ (or $j\in J_+$) exists, having a neighbor $i\in N_j$ beyond $J_-$ (respectively, beyond $J_+$). At the instant $t$ when oscillator $i$ fires $\theta_i(t)=2\pi$
and thus $\theta_j(t)\not\in\{0;2\pi\}$ since otherwise $\theta_i(t)$ would also be an endpoint.
Thus either $L(t+)\subsetneq L(t)$ and $d(t+)<d(t)$, or $\theta_j(t+)$ is not an endpoint of $L(t+)$.
On each interval of length $3T/2$ all oscillators fire. Assuming that $d(T_N)=d(0)>0$, we have
$|J_-(T_N)|+|J_+(T_N)|\le |J_-(0)|+|J_+(0)|-(N-1)\le 1$ arriving thus at the contradiction. Lemma is proved.
\begin{figure}
\center
\begin{tikzpicture}[scale=0.4]
\draw (-2.5,0) -- (2.5,0);
\draw (0,-2.5) -- (0,2.5);
\draw (0,0) circle (2cm);
\draw (0,-2) [red, very thick] arc (-90:45:2cm);
\node [label=above right:$\theta_2(t);\theta_4(t);\theta_6(t)$]  at (1.414,1.414) [circle,draw,fill, minimum size=1mm] {};
\node [label=below right:$\theta_3(t);\theta_5(t)$]  at (0,-2) [circle,draw,fill, minimum size=1mm] {};
\node [label=below right:$\theta_1(t)$]  at (2,0) [circle,draw,fill, minimum size=1mm] {};
\end{tikzpicture}
\caption{Example: $L$ is drawn red, $J_-=\{3,5\}$, $J_+=\{2,4,6\}$}\label{fig.j}
\end{figure}
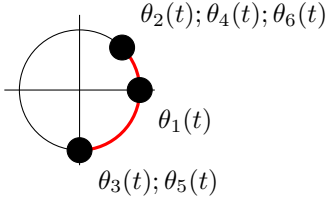
\end{IEEEproof}
\begin{corollary}\label{cor.strict-decrease}
For any constants $d_1,d_2>0$ such that $0<d_1<d_2<\pi$ there exists $\ve=\ve(d_1,d_2)>0$ that $d(\bar\theta(T_N))-d(\bar\theta(0))\le -\ve$ for any solution with $d_1\le d(\bar\theta(0))\le d_2$.
\end{corollary}
\begin{IEEEproof}
Assume, on the contrary, that a sequence of solutions $\bar\theta^{(n)}(t)$ exists such that
$d_1\le d(\bar\theta^{(n)}(0))\le d_2$, however $d(\bar\theta^{(n)}(T_N))-d(\bar\theta^{(n)}(0))\ge -1/n$. Since the set $\{\bar\theta\in [0;2\pi]^N:d_1\le d(\bar\theta)\le d_2\}$ is compact, one may assume, without loss of generality, that
the limit $\bar\theta_0\dfb\lim_{n\to\infty}\bar\theta^{(n)}(0)$ exists. Consider the solution $\bar\theta(t)$
with the initial condition $\bar\theta(0)=\theta_0$. Arbitrarily close to $T_N$ there exists a time instant $t_0$, such that none of the oscillators
fires at $t_0$ and $d(\bar\theta(t_0))=d(\bar\theta(T_N))$. Thanks to Lemma~\ref{lem.contin}, one has
$\bar\theta^{(n)}(t_0)\to \bar\theta(t_0)$ as $n\to\infty$ and thus $d(\bar\theta(T_N))=d(\bar\theta(t_0))\ge d(\bar\theta(0))\ge d_1>0$, arriving thus at a contradiction with Lemma~\ref{lem.shrink}.
\end{IEEEproof}

The proof of Theorem~\ref{thm.fix} is now immediate.
\begin{IEEEproof}[Proof of Theorem~\ref{thm.fix}]
Since the diameter is non-increasing, the limit $d_1\dfb\lim_{t\to\infty}d(\bar\theta(t))$ exists. It suffices to prove that $d_1=0$. Suppose, on the contrary, that $d_1>0$.
Denoting $d_2\dfb d(\bar\theta(0))$, one has $d_1\le d(\bar\theta(t))\le d_2$ for any $t$ due to Theorem~\ref{thm.exist}. Corollary~\ref{cor.strict-decrease} implies that
$0\le d(\bar\theta(kT_N))\le d_2-k\ve$ for any $k\ge 1$, where $\ve>0$ is constant, arriving at a contradiction. Hence the oscillators synchronize~\eqref{eq.sync}.
\end{IEEEproof}

\section{Numerical simulations}

In this section, we confirm the result of Theorem~\ref{thm.fix} by a numerical test.
We simulate a network of $N=4$ identical oscillators, whose natural frequency is $\omega=1$rad/s (and the period $T=2\pi$ s), starting with phases
$\theta_1=\pi/2$, $\theta_2=0.3\pi$, $\theta_3=0.03\pi$ and $\theta_4=0.9\pi$, thus $d(\bar\theta(0))=0.87\pi<\pi$.

\begin{figure}[h]
\center
\begin{tikzpicture}[->,>=stealth',shorten >=1pt,auto,node distance=3cm,
thick, scale=0.5,main node/.style={circle,fill=gray!20,draw,font=\sffamily\Large\bfseries,scale=0.5}]

\node[main node] (1) {1};
\node[main node] (2) [below right of=1,node distance=3cm] {2};
\node[main node] (3) [above right of=2, node distance=3cm] {3};
\node[main node] (4) [below  of=2, node distance=3cm] {4};

\path[every node/.style={font=\sffamily\small}]
(1) edge node [left] {$e_1$} (2)
(2) edge [bend left] node [left] {$e_2$} (3)
(3) edge [bend left] node [right] {$e_3$} (2)
(2) edge [bend left] node [right] {$e_4$} (4)
(4) edge [bend left] node [left] {$e_5$} (2);
\end{tikzpicture}
\caption{The network topology}\label{fig.graph}
\end{figure}
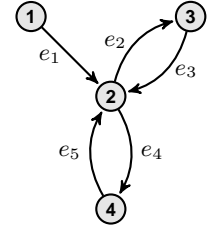

We have simulated the dynamics of the oscillators under the interaction graph, shown in Fig.~\ref{fig.graph}.
Notice that the graph in Fig.\ref{fig.graph} is rooted but \emph{not} strongly connected because the phase of the ``leading'' oscillator $1$ is unaffected by the others.

Two numerical tests have been carried out. 

\begin{figure}[h]
\center
\begin{subfigure}[b]{0.49\columnwidth}
\includegraphics[width=\columnwidth]{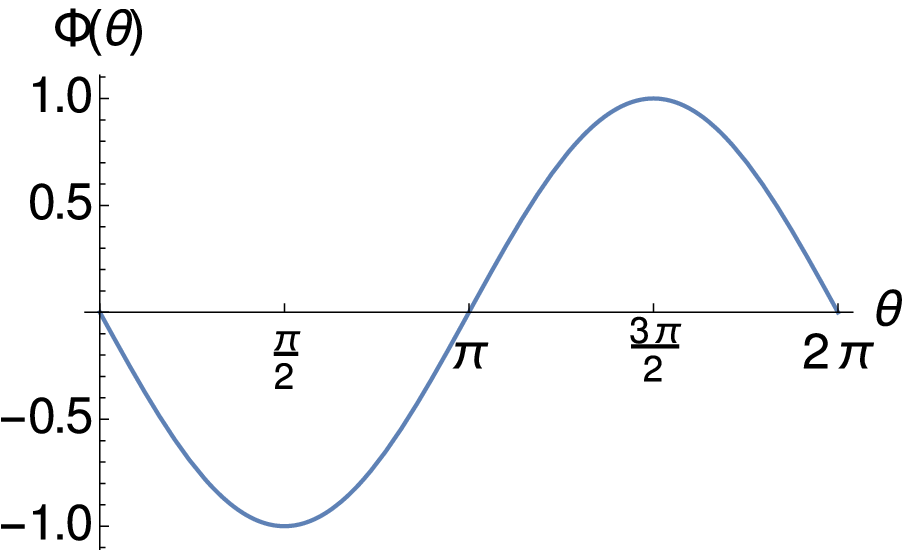}
\caption{$\Phi(\theta)=-\sin\theta$}
\end{subfigure}
\begin{subfigure}[b]{0.49\columnwidth}
\includegraphics[width=\columnwidth]{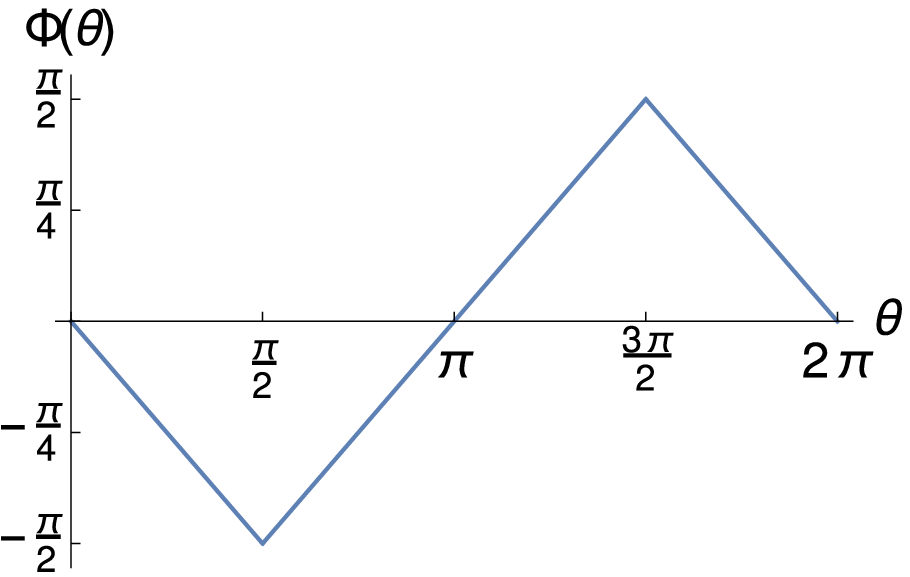}
\caption{Piecewise-linear PRC~\eqref{eq.piecewise-lin}}
\end{subfigure}
\caption{Two delay-advanced PRC maps}\label{fig.fi}
\end{figure}

\textbf{Test 1} deals with identical oscillators, having 
the delay-advanced PRC $\Phi(\theta)=-\sin\theta$ (Fig.\ref{fig.fi}a) and the gain $c=0.4$. 

\textbf{Test 2} deals with a heterogeneous network, where oscillators 2-3 have identical PRC maps
$\Phi_2(\theta)=\Phi_3(\theta)=\Phi_4(\theta)=-\sin\theta$ yet different gains $c_2=0.4$, $c_3=0.5$, $c_4=0.6$.
Furthermore, the leading oscillator $1$ has the gain $c_1=0.6$ and the following piecewise-linear PRC map (Fig.~\ref{fig.fi}b)
\be\label{eq.piecewise-lin}
\Phi_1(\theta)=\begin{cases}
-\theta,\theta\in [0;\pi/2)\\
\theta-\pi,\theta\in [\pi/2;3\pi/2]\\
2\pi-\theta,\theta\in (3\pi/2;2\pi].
\end{cases}
\ee

\begin{figure}[h]
\center
\begin{subfigure}[b]{0.49\columnwidth}
{\includegraphics[width=\columnwidth]{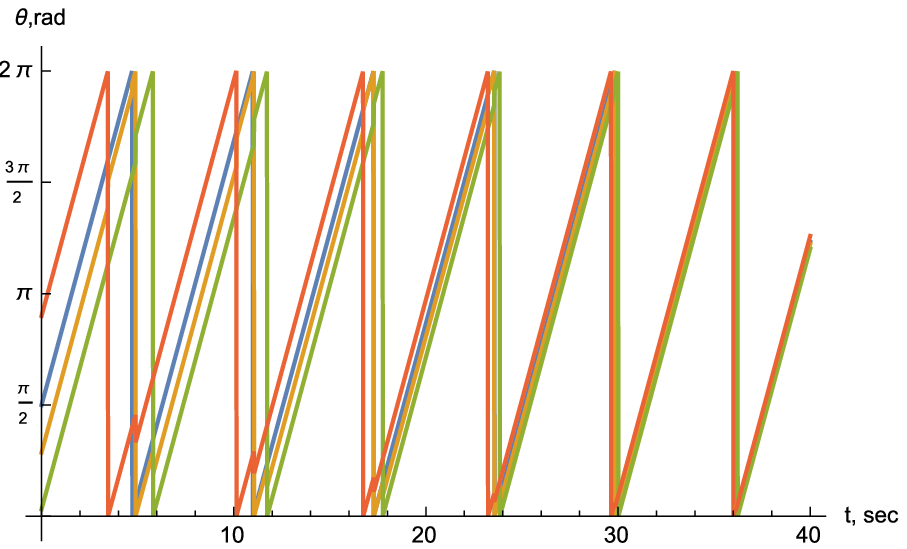}}
\caption{Test 1}
\end{subfigure}
\begin{subfigure}[b]{0.49\columnwidth}
{\includegraphics[width=\columnwidth]{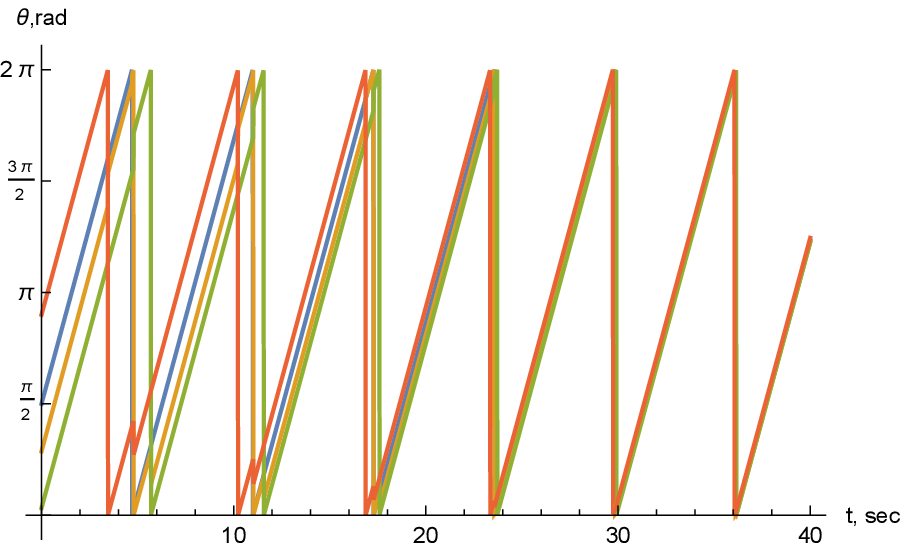}}
\caption{Test 2}
\end{subfigure}
\caption{Dynamics of the phases $\theta_i(t)$}\label{fig.phase}
\end{figure}

\begin{figure}[h]
\center
\begin{subfigure}[b]{0.49\columnwidth}
{\includegraphics[width=\columnwidth]{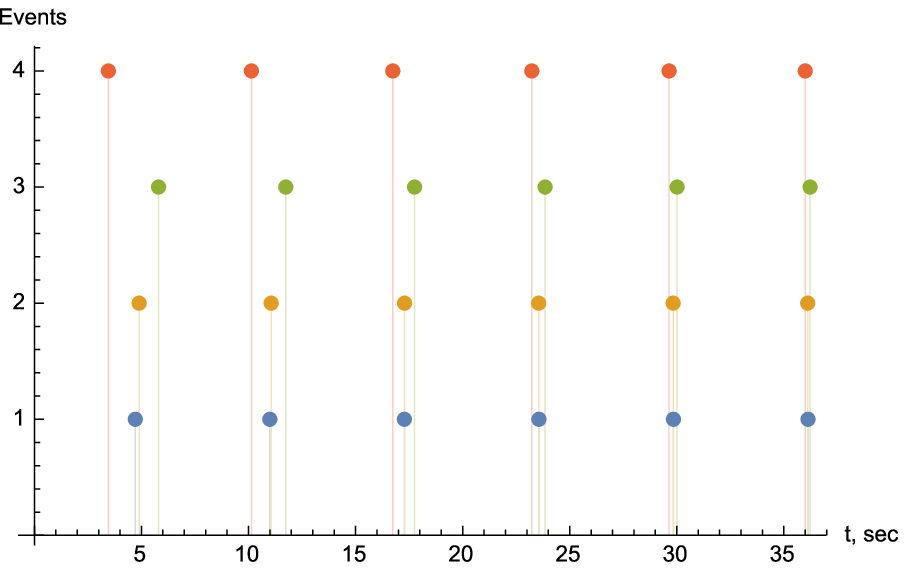}}
\caption{Test 1}
\end{subfigure}
\begin{subfigure}[b]{0.49\columnwidth}
{\includegraphics[width=\columnwidth]{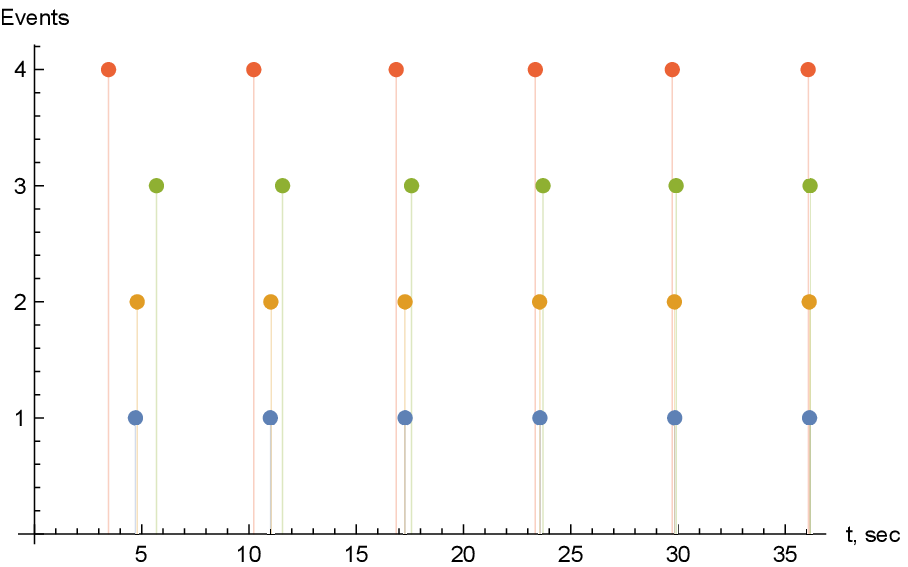}}
\caption{Test 2}
\end{subfigure}
\caption{The diagrams of events.}\label{fig.event}
\end{figure}

\begin{figure}[h]
\center
\begin{subfigure}[b]{0.24\columnwidth}
{\includegraphics[width=\columnwidth,height=\columnwidth]{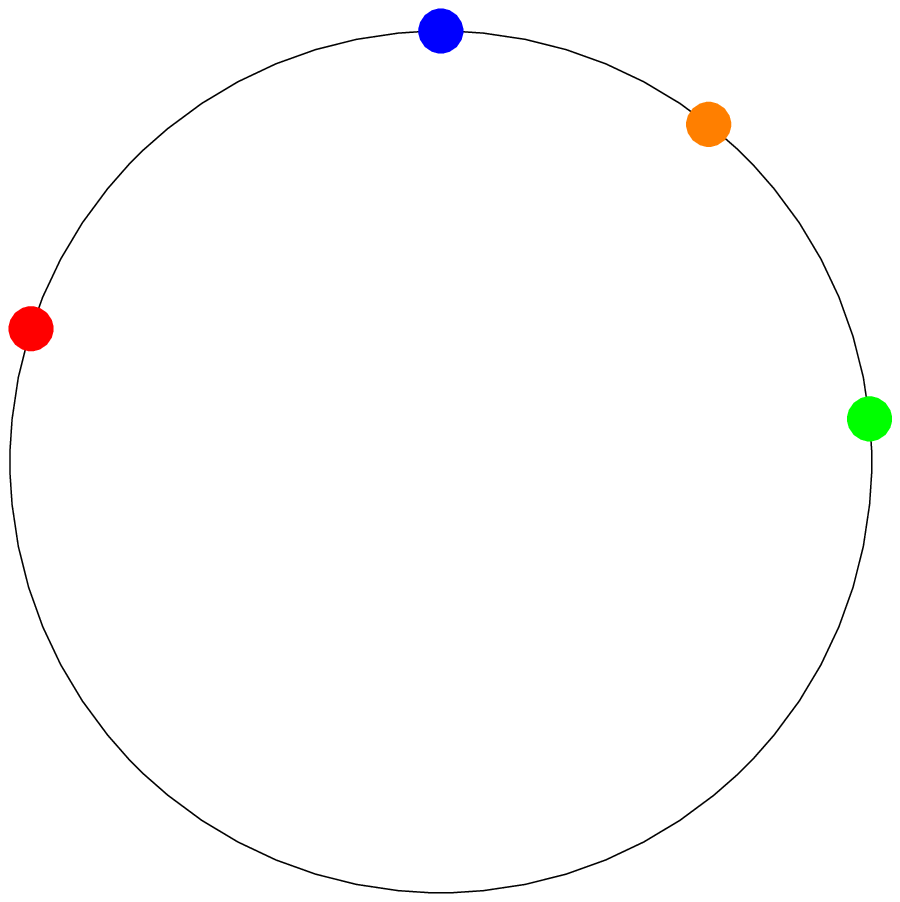}}
\caption{$t=0$s}
\end{subfigure}
\begin{subfigure}[b]{0.24\columnwidth}
{\includegraphics[width=\columnwidth,height=\columnwidth]{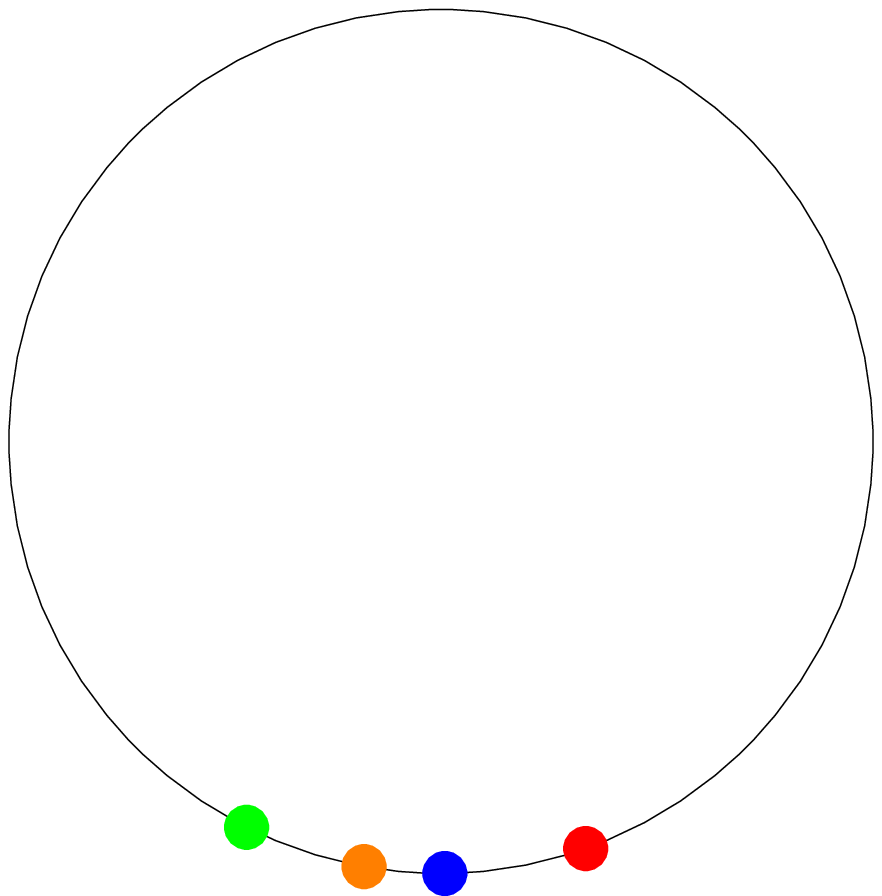}}
\caption{$t=22$s}
\end{subfigure}
\begin{subfigure}[b]{0.24\columnwidth}
{\includegraphics[width=\columnwidth,height=\columnwidth]{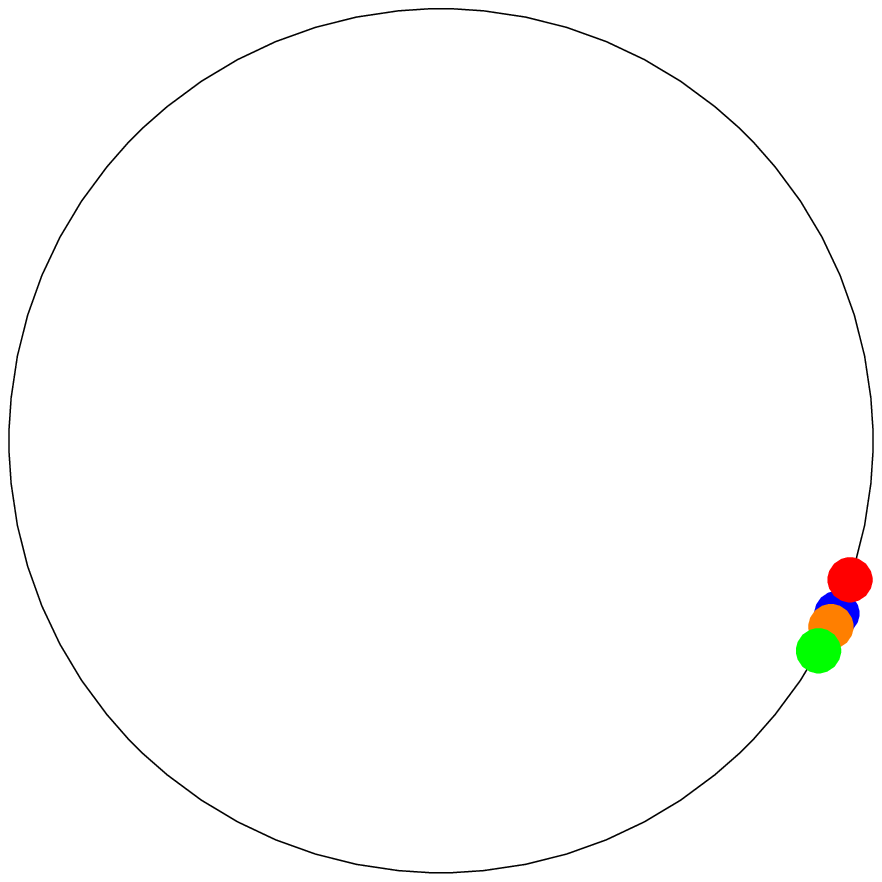}}
\caption{$t=42$s}
\end{subfigure}
\begin{subfigure}[b]{0.24\columnwidth}
{\includegraphics[width=\columnwidth,height=\columnwidth]{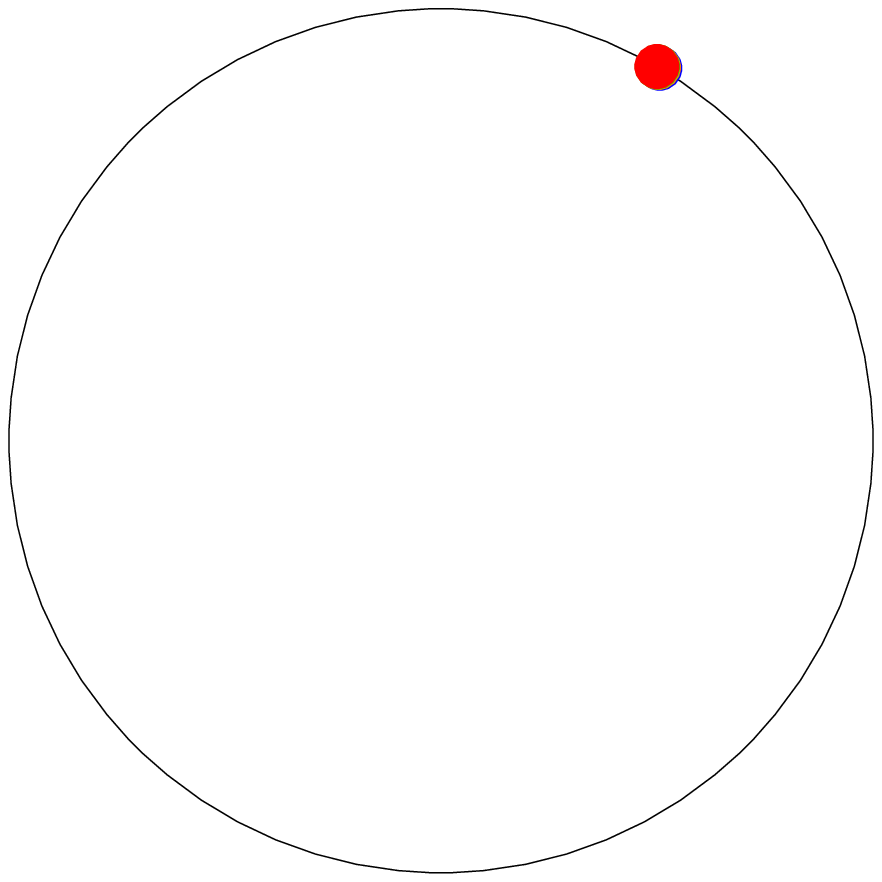}}
\caption{$t=100$s}
\end{subfigure}
\begin{subfigure}[b]{0.24\columnwidth}
{\includegraphics[width=\columnwidth,height=\columnwidth]{c0.eps}}
\caption{$t=0$s}
\end{subfigure}
\begin{subfigure}[b]{0.24\columnwidth}
{\includegraphics[width=\columnwidth,height=\columnwidth]{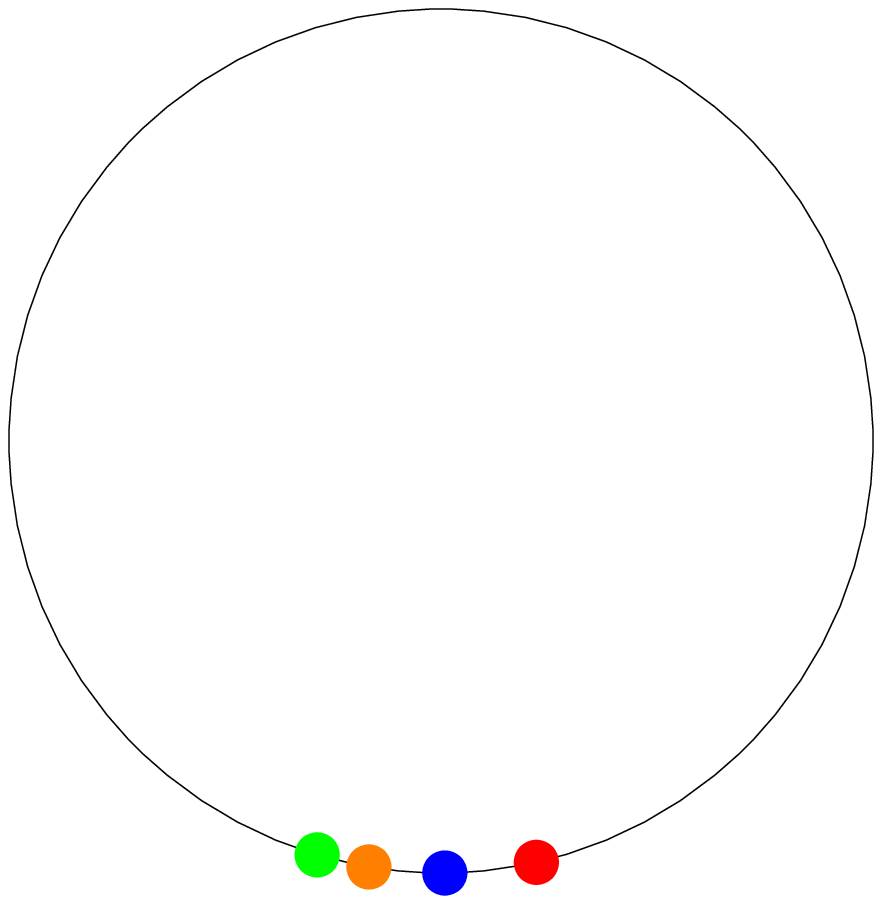}}
\caption{$t=22$s}
\end{subfigure}
\begin{subfigure}[b]{0.24\columnwidth}
{\includegraphics[width=\columnwidth,height=\columnwidth]{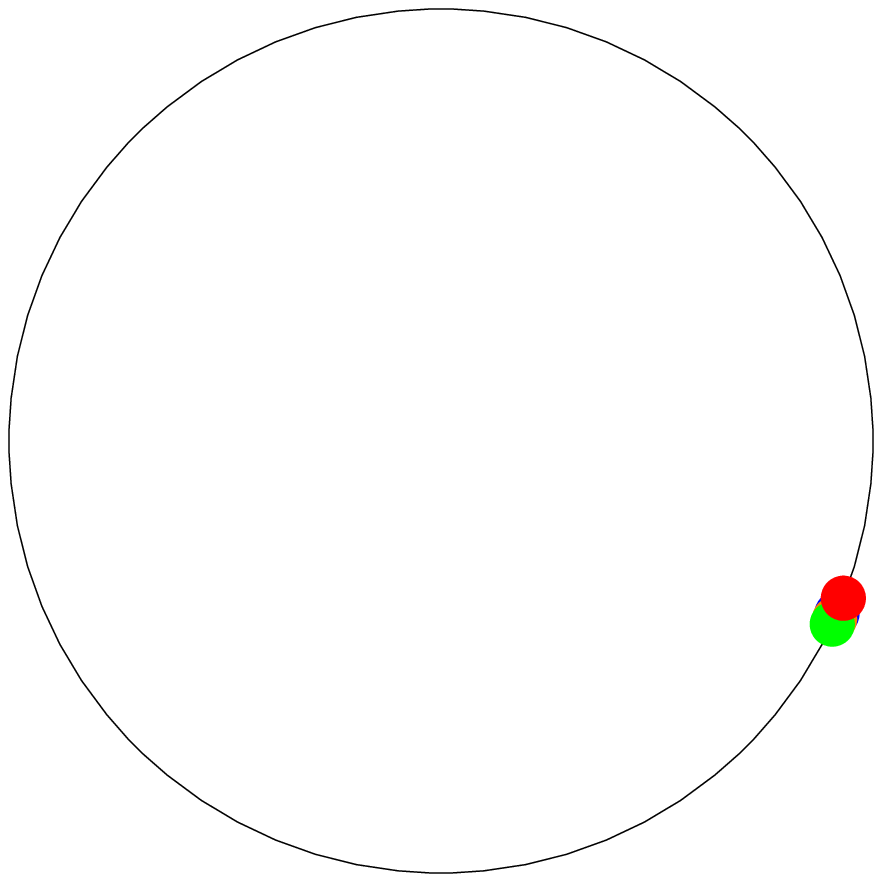}}
\caption{$t=42$s}
\end{subfigure}
\begin{subfigure}[b]{0.24\columnwidth}
{\includegraphics[width=\columnwidth,height=\columnwidth]{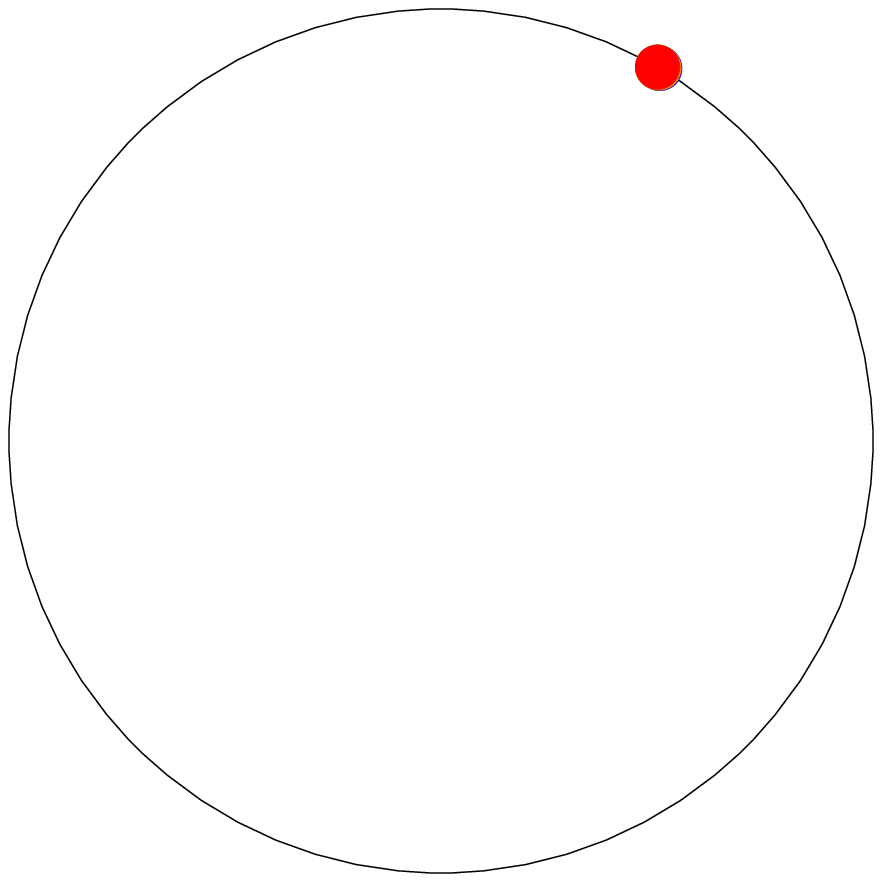}}
\caption{$t=100$s}
\end{subfigure}
\caption{Phases on $S^1$ at four time instants: the plots on top are for Test 1 and the plots on the bottom are for Test 2.}\label{fig.circle}
\end{figure}

In both numerical examples the oscillators synchronize, i.e.~\eqref{eq.sync} holds.
The corresponding dynamics of oscillators' phases
$\theta_1$ (blue), $\theta_2$ (orange), $\theta_3$ (green) and $\theta_4$ (red) are shown in Fig.~\ref{fig.phase}. Fig.~\ref{fig.event} illustrates the corresponding event diagrams:
the point $(t,i)$ on the plot in Fig.~\ref{fig.event} (where $t\ge 0$ and $i\in 1:4$) indicates that the $i$th oscillator fires an event at time $t$.
Finally, Fig.~\ref{fig.circle} illustrates synchronization of phases on the unit circle $S^1$: plots (a)-(d) correspond to Test~1, and
(e)-(h) illustrate the solutions obtained in Test~2.

\section{Conclusions and future works}\label{sec.concl}

In this paper, we have examined the dynamics of networks of pulse-coupled oscillators of the \emph{delay-advance} type. The models, studied in this paper,
describe some biological networks~\cite{GoelErmentrout:2002,IzhikevichBook} and naturally arise in problems of synchronization of networked clocks~\cite{WangDoyle:2012,WangNunezDoyle:2012}.
We have proved that the oscillators get synchronized if the maximal distance between the initial phases is less than $\pi$ and the interaction graph is static and rooted (has a directed spanning tree), which is the \emph{minimal} possible connectivity assumption. An extension to time-varying \emph{repeatedly} rooted graphs is also possible.

An important problem, which is beyond the scope of this paper and remains open even for strongly connected graphs, is synchronization under \emph{general} initial conditions.
The existing results deal mainly with all-to-all or cyclic graphs \cite{MirolloStrogatz:1990,DrorCanavier:1999,GoelErmentrout:2002,LuckenYanchuk:2012,WangNunezDoyle:2015}
which guarantee some ordering of the oscillators' events and global contraction of the return map. For instance, as was noticed in~\cite{WangNunezDoyle:2015-1}, for the PRC map
\eqref{eq.prc-lin}, the coupling gain $0.5\le c\le 1$ and the complete interaction graph, the diameter of ensemble becomes less than $\pi$ after the first event \emph{independent} of the initial condition. Another result, reported in~\cite{WangNunezDoyle:2015-1}, ensures synchronization over ``strongly rooted'' (star-shaped) and connected bidirectional graphs.
However, as noticed in Remark~\ref{rem.no-synchronize}, in general the phases of pulse-coupled oscillators do not synchronize and can e.g. split into several clusters~\cite{LuckenYanchuk:2012}; similar effects may occur due to communication delays and negative (repulsive) couplings~\cite{MalladaTang:2013}.
Even more complicated is the problem of synchronization between oscillators of different periods. One of the first results in this direction has been obtained in the recent paper~\cite{NunezWangTeelDoyle:2016}.

\bibliographystyle{IEEEtran}
\bibliography{consensus,oscillators,event}

\appendices
\section{Proof of Proposition~\ref{prop.uniqueness}}

Let $\bar\theta(0)\dfb\xi$ and $\theta_i(t)\dfb \xi_i^+ + t\omega\,\forall t\in (0;\delta_0)\,\forall i$. We are going to show that $\bar\theta(t)$ is a solution to the system~\eqref{eq.freq},~\eqref{eq.shift} on $[0;\delta_0)$ with the initial condition $\bar\theta(0)=\xi$. Indeed, on $(0;\delta_0)$ one has $\theta_i(t)<2\pi\,\forall i$, therefore,
$I(\bar\theta(t))=\emptyset$ and~\eqref{eq.freq} holds. If $I(\bar\xi)=\emptyset$, one has $\xi^+=\xi$ and hence~\eqref{eq.freq} holds also for $t=0$. Otherwise, at $t=0$
the function $\bar\theta$ jumps in accordance with~\eqref{eq.shift}: $\bar\theta(0+)=\bar\xi^+=\bar\Psi(\bar\theta(0))$.

To prove the uniqueness, notice that for arbitrary solution with $\bar\theta(0)=\bar\xi$, defined on $\Delta_0$, one has $\bar\theta(0+)=\xi^+$. Indeed, if $I(\bar\xi)=\emptyset$ then $\xi^+=\xi=\bar\theta(0)=\bar\theta(0+)$, otherwise
$\bar\theta(0+)=\xi^+$ due to~\eqref{eq.shift}. Notice now that on $(0;\delta_0)$ no oscillator can fire.
Indeed, were some events fired on this interval, the \emph{first} event instant $\tau\in(0;\delta_0)$ would be well
defined due to condition~1) in~Definition~\ref{def.solution}. Since~\eqref{eq.freq} holds on $(0;\tau)$,  $\theta_i(\tau)=\xi_i^+ + \tau\omega<2\pi\,\forall i$, arriving thus at the contradiction with the definition of $\tau$.
Therefore,~\eqref{eq.freq} holds on $(0;\delta_0)$ and $\theta_i(t)\dfb \xi_i^+ + t\omega\,\forall t\in (0;\delta_0)\,\forall i$, which ends the proof of uniqueness. $\blacksquare$

\section{Proof of Proposition~\ref{prop.positive-time}}

In the case where $\delta>d_*$ oscillator $i$ with $\theta_i(0)<2\pi-\delta$ one can take $\tau\dfb\min(T/2;\om^{-1}(\delta-d_*))$: if
$\theta_i(0)<\pi$, oscillator $i$ cannot fire earlier than at $t=T/2$ due to statement~2 of Theorem~\ref{thm.exist}, otherwise the initial phases of all oscillators belong to
$[\theta_i(0)-d_*;\theta_i(0)+d_*]\subseteq [0;2\pi-(\delta-d_*)]$ and hence no event is fired on $[0;\tau)$. We assume thus that $0<\delta\le d_*$.

We first prove the following weaker statement via induction on $m\ge 1$.
For any $d_*<\pi$ and $\delta\le d_*$ there exists $\tau_m=\tau_m(d_*,\delta)>0$ such that if $\theta_i(0)\le 2\pi-\delta$ and $d(\bar\theta(0))<d_*$, then
either oscillator $i$ does not fire on $[0;\tau_m)$, unless before its event at least $m$ other events are fired.
For $m=1$ the claim is obvious: if no event is fired, oscillator $i$ fires no earlier than at $\tau_0=\om^{-1}\delta$.
Suppose that $m\ge 2$ and the claim has been proved for $m-1$. Let $\vp_m(\delta,d_*)\dfb\min\{2\pi-\Psi^k(\theta):\theta\in [2\pi-d_*;2\pi-\delta/2],\,1\le k\le m\}>0$.
Then one can put $\tau_m\dfb \min(T/2,\om^{-1}\delta/2,\tau_{m-1}(\vp_m(\delta)))$. Consider the instant $t_0$ of the first event. At this time
one either has $\theta_i(t_0)\le d_*$ (and thus $t_{i1}>t_0+T/2$) or $\theta_i(t_0)\ge 2\pi-d_*$. In the latter case, there are two possibilities:
$\theta_i(t_0)\ge 2\pi-\delta/2$ or $\theta_i(t_0)\in [2\pi-d_*; 2\pi-\delta/2]$. The first of these possibilities implies that $t_{i1}\ge t_0\ge \om^{-1}\delta/2$,
and the second one implies that $\theta_i(t_0+)\le2\pi-\vp_m(\delta,d_*)$. Since on $[t_0;t_{i1})$ less than $m$ events are fired, one has $t_{i1}\ge t_0+\tau_{m-1}\big(\vp_m(\delta,d_*)\big)\ge\tau_m$.

It remains to notice that, due to statements~5) and 2) of Theorem~\ref{thm.exist} at most $2(N-1)$ events may occur until the oscillator fires for the first time.
Thus one can put $\tau\dfb\tau_{2(N-1)}$. $\blacksquare$ 

\end{document}